\documentclass[doublecol,linenumbers]{epl2} % for 2 columns style with line numbers
\usepackage{eurosym}
\usepackage{amsfonts}
\usepackage{amssymb}
\usepackage{amsmath}
\usepackage{graphicx}

\title{Electron-doped phosphorene: A potential monolayer superconductor}

\author{D. F. Shao\inst{1} \and W. J. Lu\inst{1}\thanks{E-mail: \email{wjlu@issp.ac.cn}} \and H. Y. Lv\inst{1}  \and Y. P. Sun\inst{1,2,3}\thanks{E-mail: \email{ypsun@issp.ac.cn}}}
\shortauthor{D. F. Shao \etal}

\institute{                    
  \inst{1} Key Laboratory of Materials Physics, Institute of Solid State Physics,
Chinese Academy of Sciences, Hefei 230031, People's Republic of China\\
  \inst{2} High Magnetic Field Laboratory, Chinese Academy of Sciences, Hefei
230031, People's Republic of China\\
  \inst{3} University of Science and Technology of China, Hefei 230026,
People's Republic of China
}


\abstract{
We predict by first-principles calculations that the electron-doped
phosphorene is a potential BCS-like superconductor. The stretching modes at the Brillouin-zone center are remarkably softened by the electron-doping,  which results in  the strong electron-phonon coupling. The superconductivity can be introduced by a doped electron density ($n_{2D}$) above $1.3 \times10^{14}$  cm$^{-2}$, and may exist over the liquid helium temperature when  $n_{2D}>2.6 \times10^{14}$ cm$^{-2}$.   { The superconductivity can be significantly tuned and enhanced by applying tensile strain.}  The maximum critical temperature   {of electron doped phosphorene} is predicted to be higher than 10 K. The superconductivity   of phosphorene   will significantly broaden the applications of this novel material.}

\begin{document}
\maketitle

The two-dimensional (2D) monolayer superconductor  bears consequences for both applications and fundamental science. It can be used as the component of nanoscale superconducting devices, such as nano superconducting quantum interference devices and nano superconducting transistors \cite{De-franceschi-nat-nano-2010,Huefner-PRB2009,Delahaye-sicence2003,Sira-PRL2007,Romans-APL2010}, with the goal of achieving single-spin sensitivity for measuring and controlling. Moreover, the high-$T_c$ superconductors  with the quasi-2D layered structures, such as MgB$_2$ \cite{Nagamatsu-nature2010},  cuprate superconductors \cite{Bednorz-1986,Wu-1987}, and iron-based superconductors \cite{Kamihara-2008}, can be seen as the assembly of multiple monolayers. Geim \emph{et al.} \cite{Geim-nature-2013} proposed to  construct the high-$T_c$-superconductor-like Van der Waals heterostructures using the monolayer superconductors, which may be helpful for the exploration of new  high-$T_c$ superconductors. A graphene-like monolayer superconductor seems to be the natural choice of such applications. It is suggested that the carrier-doped graphene
\cite{Profeta-nat-phys-2012,Dai-nanoscale-2012,Si-PRL2013} and graphane \cite{Savini-prl2010} may exhibit superconductivity with notable $T_c$. However, the experimental evidences are still lacking. Using the liquid-gate method, one can introduce the superconductivity  into the few-layer semiconductor MoS$_2$ \cite{Ye-science2012}, which may be driven by the electron-phonon coupling \cite{Ge-PRB2013}. But the thickness of such material is still far from one layer.  More monolayer superconductor candidates  still need to be found.

Here we show a potential monolayer superconductor: electron-doped phosphorene. Based on the density functional theory (DFT) calculations, we found the electron-doping can make the stretching modes at zone center significantly softened, leading to a strong electron-phonon coupling.  The superconductivity starts showing up when the carrier density ($n_{2D}$) is $1.3 \times10^{14}$ cm$^{-2}$. When  $n_{2D}>2.6 \times10^{14}$ cm$^{-2}$, the $T_c$  exceeds  the liquid helium temperature.   {Moreover, the application of tensile strain can significantly tune and enhance the superconductivity.} The maximum $T_c$ is predicted to be higher than 10 K. Our prediction can be readily verified by  the liquid-gate method  \cite{Ye-science2012,Ueno-Nat-Mater-2008,Ueno-Nat-nano-2011,Ye-nat-mater-2010},  or the adsorption of the alkali/alkaline-earth metal atoms.

\begin{figure}
\includegraphics[width=0.95\columnwidth]{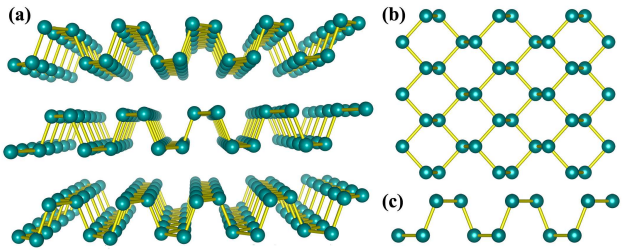}\caption{\label{fig1_structure} (a) Structure of bulk black phosphorus.  (b) and (c) are the   top and side views of the     structure of phosphorene, respectively.}
\end{figure}

Figure \ref{fig1_structure} shows the structures of bulk black phosphorus (black-P) and monolayer black-P (which is called phosphorene). Phosphorene  can be obtained by exfoliating the  black-P \cite{Li-natnano-2014,Lu-arxiv}, the most stable allotropic form of phosphorus and the only  layered structure of an elemental solid besides graphitic carbon  \cite{Zhu-arxiv}. The phosphorus atoms are covalently bonded, forming a special puckered 2D structure  \cite{Brown-Acta1965,Slater-PR1962,Cartz-jcp1979}. Both bulk black-P and phosphorene are direct-gap semiconductors, in which the gap increases from 0.3 eV
in bulk black-P to 2 eV in phosphorene with decreasing the number of layers \cite{Takao-physica1981,Keyes-PR1953,Warschauer-JAP1963,Maruyama-physica1981,Asahina-JPSJ1982,Akahama-JPSJ1983,Appalakondaiah-PRB2012}. Recent theoretical and experimental works showed that phosphorene is a good candidate to realize the novel optoelectronic, electronic, thermoelectric, and nano-mechanical devices because of such band properties \cite{Lu-arxiv,Li-natnano-2014,Koenig-arxiv,Liu-arxiv,Qiao-arxiv,Xia-arxiv,Lv-thermoelectric}. If the superconductivity can be introduced, the applications of phosphorene  will be significantly broadened.

In this letter, we  performed the  DFT calculations to investigate the potential superconductivity in electron-doped phosphorene. Calculations were carried out using   {the} QUANTUM-ESPRESSO package \cite{Giannozzi-jpcm2009}. The ultrasoft pseudopotential within the generalized
gradient approximation  (GGA) according to the Perdew-Burke-Ernzerhof  \cite{PBE} was used. The energy cutoff for the plane-wave basis set was 40 Ry. The Brillouin
zone was sampled with a $32\times24\times1$ mesh of $k$ points. The Vanderbilt-Marzari Fermi smearing method with a  smearing parameter of $\sigma=0.02$ Ry  was used for the calculations of the total energy and electron charge density.   Phonon spectra and electron-phonon coupling constants were calculated using
density-functional perturbation theory \cite{DFPT} with an $8\times8\times1$ mesh of $q$ points. The double
Fermi-surface averages of electron-phonon matrix elements were calculated
using the tetrahedron method on grids of $160\times120\times1$ $k$
points. To simulate the monolayer, a vacuum layer more than 10 {\AA} was introduced.\footnote{ For the low doping concentration of $x=0.1$ electrons/cell, the smaller $\sigma=0.005$ Ry and the denser  $64\times48\times1$ mesh of $k$ points were used. For the vacuum layer, we tested  different values of the thickness. It turns out a vacuum layer more than 10 {\AA}  can lead to the converged lattice parameters ($a=3.32$ {\AA} and $b=4.64$ {\AA}) and band structure (the band gap is $\sim0.9$ eV)  in good agreement with the previous PBE calculation \cite{Qiao-arxiv}.} Electron-doping was simulated by adding electrons
into the system, together with a compensating uniform positive background \cite{Ge-PRB2013}.   For each doping concentration, we relaxed the atomic positions with the fixed  in-plane lattice constants of the optimized lattice structure of the undoped   {phosphorene}.

\begin{figure}
\includegraphics[width=0.98\columnwidth]{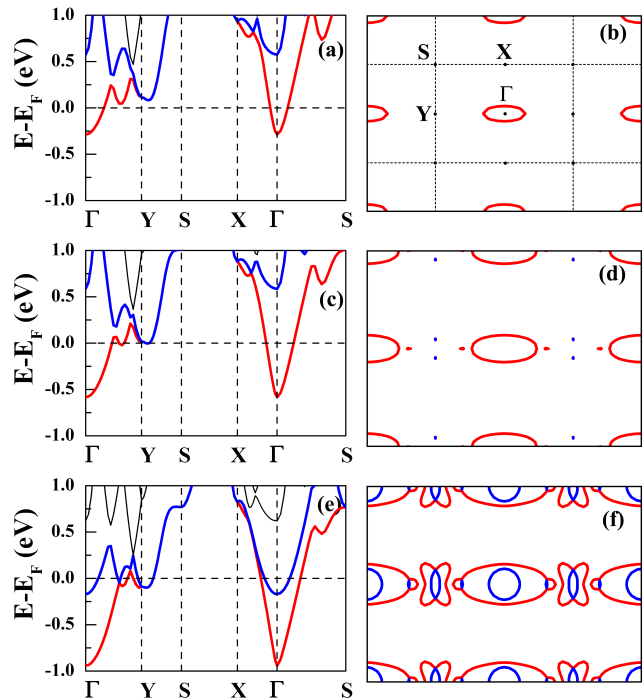}

\caption{\label{fig2-band}Conduction bands and Fermi surface of phosphorene.
(a), (c), and (e) are the band structures with respect to the doping concentrations
of $x=0.1$, 0.3, and  0.7 electrons/cell, respectively. 
(b), (d), and (f) show the corresponding Fermi sheets. The 
red and blue lines denote the Fermi sheets from the lowest and second-lowest
conduction band, respectively. The high symmetry points are denoted in (b).}

\end{figure}

Figure \ref{fig2-band} shows the conduction bands and the corresponding
Fermi sheets of three typical doping concentrations.  The effect of doping obviously enlarges the Fermi surface. For the low doping
concentrations, electrons occupy the states near the conduction band minimum at $\Gamma$
point, forming an oval Fermi sheet around the zone center, as shown
in figs. \ref{fig2-band} (a) and (b). When the doping concentration is up to   $x=0.3$ electrons/cell,
the second-lowest conduction band starts 
crossing the Fermi level ($E_{F}$) at the place near $Y$ point, and new sheets
start  to show up, as shown in figs.  \ref{fig2-band}
(c) and (d).  These sheets expand with  increasing the doping concentration. When the doping
concentration  is  above $x=0.7$ electrons/cell, the $\Gamma$ valley of the second lowest
conduction band is occupied. The valley becomes the energy minimum
of such band, and forms a circle-like Fermi sheet around the zone center
(figs. \ref{fig2-band} (e) and (f)). Such relative shift in energy
of the conduction band valleys with doping was also found in the electron-doped monolayer MoS$_{2}$ \cite{Ge-PRB2013}. 

\begin{figure}
\includegraphics[width=0.99\columnwidth]{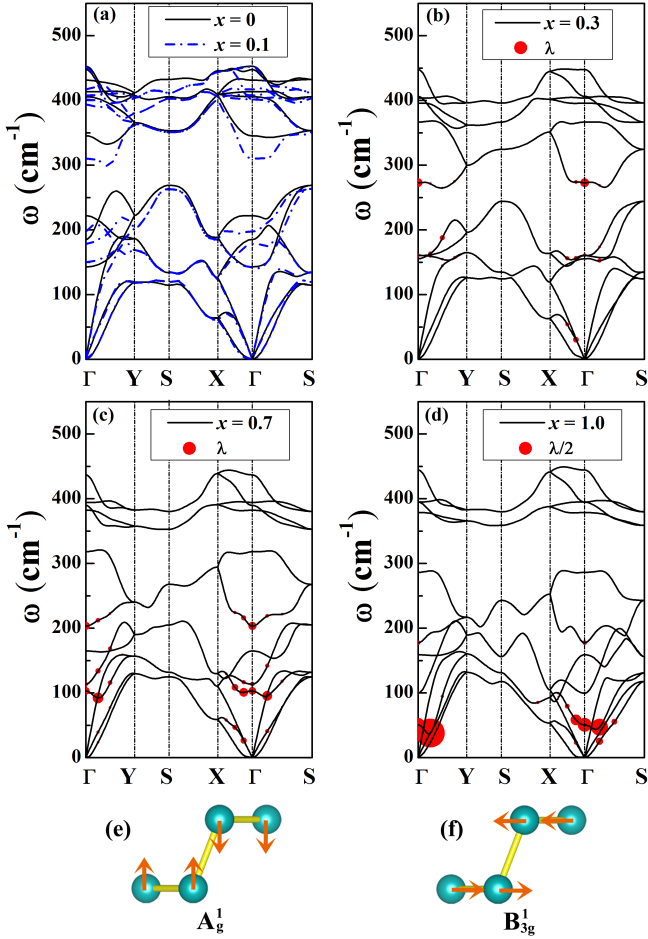}

\caption{\label{fig3_phonon} Phonon dispersions of phosphorene with some
typical doping concentrations: (a) $x=0$ (pristine sample, black solid
lines) and $x=0.1$ (blue dashed lines), (b) $x=0.3$, (c) $x=0.7$,
and (d) $x=1.0$ electron/cell. The phonon dispersions in (b), (c), and (d) are
decorated with symbols, proportional to the partial electron-phonon coupling strength $\lambda_{\nu\mathbf{q}}$.   The  $A_{g}^{1}$ (e) and $B_{3g}^{1}$  (f) stretching modes in
zone center are showed in the bottom.}
\end{figure}

The calculated phonon dispersions of the pristine and electron-doped phosphorene
are presented in fig. \ref{fig3_phonon}. For the pristine phosphorene,
our calculation is in good agreement with the previous result  reported
by Zhu and Tom{\'{a}}nek  \cite{Zhu-arxiv}.    
  {Compared with the  zone center modes of phosphorene rencently measured and bulk black phosphorus previously reported \cite{Li-natnano-2014,Kaneta-1982 , SUGAI-JPSJ1981, Kaneta-JPSJ1986-1, Kaneta-JPSJ1986-2}, our calculated zone center modes seem to have slightly smaller frequencies. It might be due to the overestimation of lattice constant for PBE \cite{Qiao-arxiv}.  However, our calculation is still enough to discuss the variation trends of the properties upon doping.} In  fig.
\ref{fig3_phonon}, one can notice that  the highest four optical branches  are slightly softened while other five optical branches are softened  significantly  upon doping. A remarkable softening can be found
in the high frequency optical zone center $A_{g}^{1}$ mode associated
with the outplane stretching (fig.  \ref{fig3_phonon} (e)). The 
region of the reciprocal space where the softening is observed matches
the diameter $2k_{F}$ of the Fermi sheets around $\Gamma$, which
is a typical signature of the Kohn effect  \cite{Kohn-PRL1959}. Such
softening generally exists in some superconductors  such as MgB$_{2}$  \cite{Kong-PRB2001,Kortus-prl2001},
hole-doped diamond  \cite{Boeri-PRL2004,Giustino-prl2007}, hole-doped graphane  \cite{Savini-prl2010},
etc.. When the doping concentration is above $x=0.7$  electrons/cell, $A_{g}^{1}$ mode
is further softened. This is due to a new Kohn anomaly caused by
the new $\Gamma$ centered circular Fermi sheets from the second-lowest
conduction band (see figs. \ref{fig2-band} (e) and (f)). Another prominent softening happens in the low frequency
optical mode $B_{3g}^{1}$ at zone center, which is related to the
inplane stretching (fig.  \ref{fig3_phonon} (f)). For the pristine
phosphorene, the $B_{3g}^{1}$ mode has the highest energy among the
low frequency optical modes at zone center. With the electron-doping,
the $B_{3g}^{1}$ mode significantly softens, and gradually becomes
frequency minimum of the optical modes at $\Gamma$.  Besides the two
remarkable softenings at the zone center, some other anomalies, such
as the anomaly at $\mathrm{\mathbf{q}}\approx\frac{1}{4}\Gamma Y$ (figs. \ref{fig3_phonon} (c) and (d)),
have been found in the low frequency optical branches. These anomalies
start to show up for $x=0.3$  electrons/cell, when the new Fermi sheets are introduced into the system. It indicates
such anomalies might be due to   
 {the nesting between the  Fermi sheets}.  When   $x\geqslant 1.2$ electrons/cell,  frequencies of $B_{3g}^{1}$ and its adjacent  modes   become negative, indicating the appearance of the lattice instability induced by doping. Such instability suggests the electron-doping limit of phosphorene should be smaller than 1.2 electrons/cell.

\begin{figure}
\includegraphics[width=0.98\columnwidth]{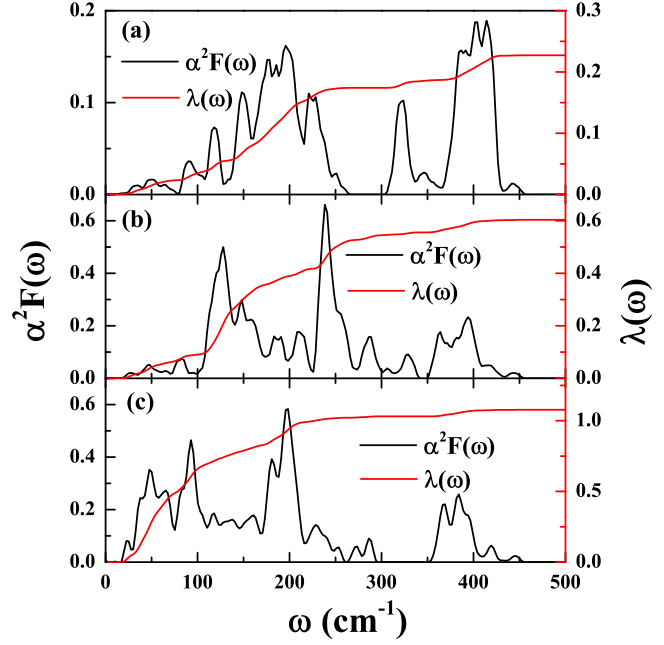}

\caption{\label{fig4_a2f}   (a)   Eliashberg function (left) and   the integrated  electron-phonon coupling strength (right) for the elelctron-doped  phosphorene with  $x=$  0.1 (a), 0.5 (b), and  1.0 (c) electron/cell. }
\end{figure}

\begin{figure}
\includegraphics[width=0.98\columnwidth]{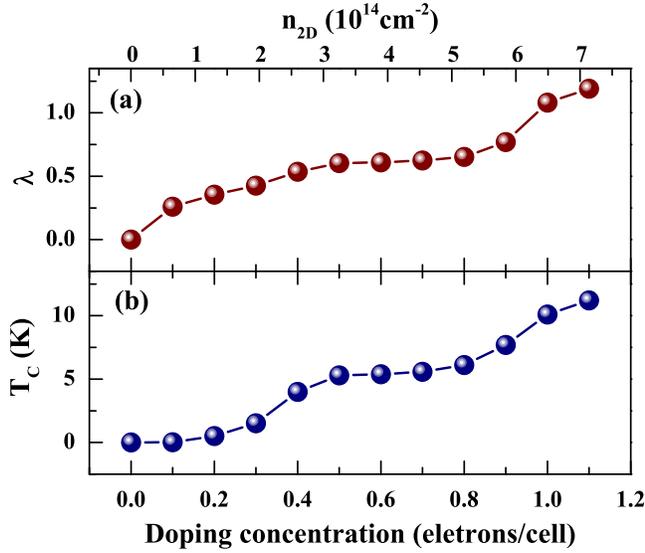}

\caption{\label{fig5_TC}   (a) Total electron phonon coupling strength
$\lambda$ and (b) predicted $T_{c}$ with different doping concentrations. The carrier density corresponding to the doping concentration is shown at the top axis.}
\end{figure}

Figure \ref{fig4_a2f}  plots the Eliashberg spectral function:

\begin{equation}
\label{eq:a2f}
\alpha^{2}F(\omega)=\frac{1}{N(E_{F})}\underset{\mathbf{k},\mathbf{q},\nu,n,m}{\sum}\delta(\epsilon_{\mathbf{k}}^{n})\delta(\epsilon_{\mathbf{k}+\mathbf{q}}^{m})\mid g_{\mathbf{k},\mathbf{k}+\mathbf{q}}^{\nu,n,m}\mid^{2}\delta(\omega-\omega_{\mathbf{q}}^{\nu}),
\end{equation}
where $N(E_{F})$ is the density of states at $E_{F}$, $\omega_{\mathbf{q}}^{\nu}$ is phonon frequency, $\epsilon_{\mathbf{k}}^{n}$
is electronic energy, and $g_{\mathbf{k},\mathbf{k}+\mathbf{q}}^{\nu,n,m}$
is electron-phonon coupling matrix element. The total electron-phonon
coupling strength  is then 
\begin{equation}
\lambda=2\int_{0}^{\infty}\frac{\alpha^{2}F(\omega)}{\omega}d\omega=\underset{\nu\mathbf{q}}{\sum}\lambda_{\nu\mathbf{q}}.\label{eq:lambda}
\end{equation}
$\lambda_{\nu\mathbf{q}}$ are visualized as red circles in fig. \ref{fig3_phonon}.
Acoustic modes near $\Gamma$ contribute a small peak
to the $\alpha^{2}F(\omega)$ (fig. \ref{fig4_a2f} (a) and (b)). But the
predominant contribution is from the softened optical modes. The large
contribution of electron-phonon coupling from the $A_{g}^{1}$ mode
 leads to a sharp peak in $\alpha^{2}F(\omega)$  at high frequency (fig. \ref{fig4_a2f})). Another
remarkable contribution is from the softening of $B_{3g}^{1}$ mode.  Moreover,
one can notice that when new Fermi sheets are introduced, the $\lambda_{\mathbf{q}}$
at some \textbf{q}-vector, such as $\mathrm{\mathbf{q}}\approx\frac{1}{4}\Gamma Y$ (see figs. \ref{fig3_phonon} (c) and (d)),
is largely enhanced.  It indicates that the   
 {intersheet nesting} plays an important role in the electron-phonon
coupling as well,  {similar to the cases in hole doped CuAlO$_2$ \cite{Katayama-Yoshida-2003,Nakanishi-2012-1}  and electron doped monolayer MoS$_2$ \cite{Ge-PRB2013}}.  For the high doping concentration, because of the very low $\omega$ in the denominator of eq. (\ref{eq:lambda}), the contributions of  $B_{3g}^{1}$ and its adjacent  modes to the electron-phonon coupling become predominant. 

We estimated the $T_{c}$ based on the Allen-Dynes formula \cite{Allen-Dynes}
\begin{equation}
T_{C}=\frac{\omega_{log}}{1.2}\exp\left(-\frac{1.04(1+\lambda)}{\lambda-\mu^{*}-0.62\lambda\mu^{*}}\right)\text{,}\label{eq:TC}
\end{equation}
 where  the Coulomb pseudopotential $\mu^{*}$ is set to a typical value of $\mu^{*}=0.1$.   The logarithmically averaged characteristic phonon frequency $\omega_{log}$
is defined as
\begin{equation}
\omega_{log}=\exp\left(\frac{2}{\lambda}\int\frac{d\omega}{\omega}\alpha^{2}F(\omega)\log\omega\right).
\end{equation} The calculated total electron-phonon coupling strength $\lambda$
and $T_{c}$ of the electron-doped phosphorene with  different doping concentrations are presented in figs.
\ref{fig5_TC} (a) and (b), respectively. For the low doping concentration, the electron-phonon
coupling is  weak. It  gradually increases with increasing doping concentration.  When the doping concentration is up to $x=0.2$ electrons/cell ($1.3 \times10^{14}$ cm$^{-2}$), the superconductivity starts showing up. The calculated $\lambda = 0.36$ leads to the superconductivity with $T_{c}\sim 0.5$ K, which already can be detected in routine measurements. When $x>0.4$ electron/cell ($n_{2D}>2.6 \times10^{14}$ cm$^{-2}$), the value of $\lambda$ is above 0.54, generating the superconductivity with $T_c$   above the liquid helium temperature of 4.2 K. The maximum $T_c$  is predicted to be 11.2 K when $x=1.1$ electrons/cell. Such value can be comparable to the maximum $T_{c}$ of    
high pressure phase of phosphorus  {\cite{Kawamura-ssc1984,Wittig-science1968,Nakanishi-2012-2}}, clearly indicating that the electron-doped
phosphorene may be another superconducting phase of phosphorus. 

 The electron doping can be processed using the liquid-gate method, which is considered as a clean route to dope carrier without introducing randomly distributed charged impurities which strongly scatter charge carriers. \cite{Ye-science2012,Ueno-Nat-Mater-2008,Ueno-Nat-nano-2011,Ye-nat-mater-2010,liquid-zno, liquid-review} Applying the method to a film with the thickness of several unit cells,   an extremely high density of electrons  can be accumulated in a very thin depth below the sample surface \cite{Ye-science2012}.  Due to the very large interlayer distance of black phosphorus ($\sim$ 5 \AA), it is reasonable to assume the electrons  doped by liquid-gate method will be mainly confined in the top monolayer \cite{Ye-science2012}. Therefore, our calculation can be seen as a direct simulation of the liquid-gate doping. The maximum $n_{2D}$ of the method reported is $8 \times10^{14}$ cm$^{-2}$ \cite{liquid-zno, liquid-review}. Since the  superconductivity  we predicted is detectable  when   $n_{2D}$  above a low carrier density of $1.3 \times10^{14}$  cm$^{-2}$, our  prediction  can be  readily verified using such method. 

In order to obtained a free standing superconductor, the chemical doping is necessary. Intercalation is an usual method to dope electrons into layered materials. However, for the black phosphorus system, to the best of our knowledge, there  are no homogeneous and stable donor intercalated compounds yet \cite{intercalation}. Although the donor-type atoms or molecules can hardly be intercalated inside the interlayers, it is suggested that those can be adsorbed at the surface \cite{Maruyama-physica1981}, which implies the chemical doping by adsorption for monolayer will be much easier and more efficient than that for bulk.  Alkali atom Cs might be a promising donor-type atom for the adsorption \cite{intercalation}.  The bands of the adsorbed atoms in phosphorene may occur at the $E_{F}$ and have multiple beneficial effects on $\lambda$ \cite{Csanyi-Natphys-2005,Profeta-nat-phys-2012}, which may enhance the superconductivity.
 
 {Since the strain is proven to be a key method to control the properties of nanoscale material \cite{Si-PRL2013,Lv-thermoelectric}, we also investigated the strain effects in the electron doped phosphorene. We found that the strain can significantly tune and enhance the superconductivity. As an example, fig. \ref{fig6_strain} shows the influence of  tensile strain along armchair direction on electron doped phosphorene with doping concentraiton of $x=0.4$  electrons/cell. The  $B_{3g}^{1}$ mode is largely softened by strain, which strongly enhances the electron phonon coupling and raises $T_{c}$. When a strain about 8\% is applied, the maximum $T_{c}=12.2$ K can be obtained. One can notice the strain strongly suppressed $\omega_{log}$, which limits the increase of $T_{c}$. Moreover, a very large strain can make the  $B_{3g}^{1}$ mode unstable. Therefore, a very high $T_{c}$ cannot be achieved in electron doped phosphorene. However, the application of strain is still an useful method to achieve superconductity with $T_{c}$ over 10 K  for  low doping concentration.

\begin{figure}
\includegraphics[width=0.98\columnwidth]{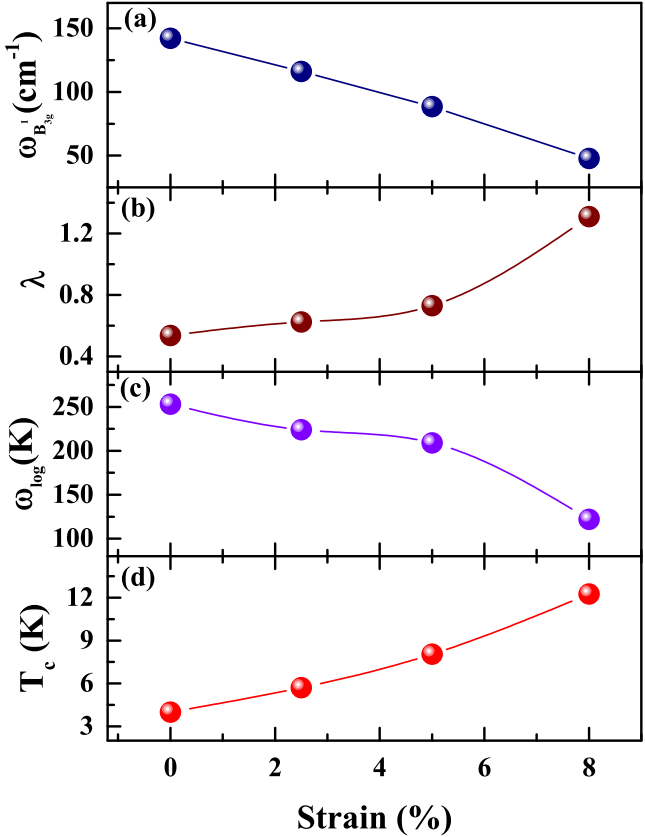}

\caption{\label{fig6_strain}  The effect of strain  along armchair directrion for the 0.4 electron doped phosphorene: The variation  of (a) frequencis of $B_{3g}^{1}$, (b) total electron phonon coupling strength
$\lambda$, (c) the logarithmically averaged characteristic phonon frequency $\omega_{log}$ and  (d) predicted $T_{c}$ under different strains along armchair direction. }
\end{figure}}

In conclusion, we have carried out the first-principles study of phosphorene
and found that the phonon mediated superconductivity   can be introduced by a doped electron density above $1.3 \times10^{14}$  cm$^{-2}$. When  $n_{2D}>2.6 \times10^{14}$ cm$^{-2}$, the $T_c$ exceeds the liquid helium temperature.  {Moreover, the superconductivity can be significantly tuned and enhanced by strain.} A maximum $T_{c}$ over 10 K can be achieved.  The electron-phonon
coupling is majorly contributed by the softening of the outplane stretching
mode $A_{g}^{1}$ and the inplane stretching mode $B_{3g}^{1}$. Our prediction
indicates that phosphorene can be a good platform to realize the nanoscale superconducting devices. Moreover, it can be
the ``atomic-scale Lego'' to construct the van der waals heterostructure
for the attempt of searching for the high $T_{c}$ superconductivity.

\acknowledgments
This work was supported by the National Key Basic Research under Contract No. 2011CBA00111, the National Nature Science Foundation of China under Contract Nos. 11304320 and 11274311, the Joint Funds of the National Natural Science Foundation of China and the Chinese Academy of Sciences’ Large-scale Scientific Facility (Grand No. U1232139), and Anhui Provincial Natural Science Foundation under Contract No. 1408085MA11. The calculation was partially performed at the Center for Computational Science, CASHIPS.

\end{document}